\documentclass[11pt,a4paper]{article}

\usepackage{tikz,pgfplots}
\usetikzlibrary{positioning}
\usepackage{pgffor}

\usepackage{xcolor}
\usepackage{soul}

\usepackage{authblk}
\title{Deep learning velocity signals allows to quantify turbulence intensity}

\author[1]{Alessandro Corbetta}
\author[2]{Vlado Menkovski}
\author[3]{Roberto Benzi}
\author[4]{Federico Toschi\footnote{Corresponding author. Email: f.toschi@tue.nl.}}
\affil[1,4]{Department of Applied Physics, Eindhoven University of
  Technology, The Netherlands}
\affil[2,4]{Department of Mathematics and Computer Science, Eindhoven University of
  Technology, The Netherlands}
\affil[3]{Department of Physics, University of Rome Tor Vergata,
  Italy}
\affil[4]{CNR-IAC, Rome, Italy}

\date{}
\usepackage{graphicx}
\usepackage{tikz}
\usetikzlibrary{positioning,shadows,fadings}

\usepackage{pgf}
\usepackage{pgfplots}

\tikzset{%
  every neuron/.style={
        circle,
    draw,
    minimum size=1cm
  },
  neuron missing/.style={
    draw=none, 
    scale=4,
    text height=0.333cm,
    execute at begin node=\color{black}$\vdots$
  },  
}

\begin{document}

\maketitle

\begin{abstract}  
  Turbulence, the ubiquitous and chaotic state of fluid motions, is
  characterized by strong and statistically non-trivial fluctuations
  of the velocity field, over a wide range of length- and time-scales,
  and it can be quantitatively described only in terms of statistical
  averages. Strong non-stationarities hinder the possibility to
  achieve statistical convergence, making it impossible to define the
  turbulence intensity and, in particular, its basic dimensionless
  estimator, the Reynolds number.

  Here we show that by employing Deep Neural Networks (DNN) we can
  accurately estimate the Reynolds number within $15\%$ accuracy, from
  a statistical sample as small as two large-scale eddy-turnover
  times. In contrast, physics-based statistical estimators are limited
  by the rate of convergence of the central limit theorem, and
  provide, for the same statistical sample, an error at least $100$
  times larger. Our findings open up new perspectives in the
  possibility to quantitatively define and, therefore, study highly
  non-stationary turbulent flows as ordinarily found in nature as well
  as in industrial processes.
\end{abstract}



Turbulence is characterized by complex statistics of velocity
fluctuations correlated over a wide range of temporal- and
spatial-scales.  These range from the integral scale, $L$,
characteristic of the energy injection (with correlation time $T_L$),
to the dissipative scale, $\eta \ll L$, characteristic of the energy
dissipation due to viscosity (with correlation time
$\tau_\eta \ll T_L$). The intensity of turbulence directly correlates
with the width of this range of scales, $L/\eta$ or $T_L/\tau_\eta$,
commonly dubbed inertial range.

In statistically stationary, homogeneous and isotropic turbulence
(HIT), the width of the inertial range, is well known to correlate
with the Reynolds number, $Re$, defined as $Re=v_{rms} L/ \nu$, where
$v_{rms}$ is the characteristic velocity fluctuation at the integral
scale, and $\nu$ is the kinematic viscosity. Therefore the value of
the Reynolds number is customarily used to quantify turbulence
intensity.  While this value remains well-defined in laboratory
experiments, performed under stationary conditions, and for fixed flow
configurations ($L=const$), its quantification results impossible when
we consider turbulence in open environments (as in many outstanding
geophysical situations) or in non-stationary situations (such as
turbulent/non-turbulent interfaces).  This observation is linked to
the question: can we estimate turbulence intensity from fluctuating
velocity signals of arbitrary (short) length?  For statistically
stationary conditions, this is indeed possible provided enough
statistical samples are available and by using appropriate
physics-based statistical averages of the fluctuating velocity field.
For non-stationary turbulent flows (i.e.  changing on timescales
comparable with the large-scale correlation times) the question itself
appears meaningless. In these conditions, the intertwined complexity
of a slow large-scale dynamics and of fast, but highly intermittent,
small-scale fluctuations, makes impossible to estimate reliably the
width of the inertial range.

In this paper we demonstrate, using a proof of concept, that our
fundamental question can be answered by a suitable use of machine
learning. We propose a machine learning Deep Neural Network (DNN)
model capable of estimating turbulence intensity within $15\%$
accuracy from short velocity signals (duration $T$: approximately two
large-scale eddy turnover times, i.e. $T\approx 2\,T_L$, where
$T_L\approx L/v_{rms}$). We remark that analyzing the same data via
standard statistical observables of turbulence leads to
quantitatively meaningless results (predictions between
$10^{-2}$ and $10^{2}$ times the true value).


We train the DNN model to predict turbulence intensity using 
(short) Lagrangian velocity signals obtained from HIT. As Lagrangian
velocities are one of the most intermittent features of turbulence, we
are choosing the most difficult case for our proof of concept. The
Lagrangian velocity signals, $v(t)$, that we employ are obtained as
the superposition of different strongly chaotic time signals,
$u_n(t)$, derived from a shell model of
turbulence~\cite{biferale2003shell,l1998improved} (see Methods).  The
velocity signals, $v(t)$, are known to match the statistical
properties of the velocity experienced by a passive Lagrangian
particle in HIT~\cite{boffetta2002lagrangian}. The shell model
describes the nonlinear energy transfer among different spatial
scales, $l_n=1/k_n$, where $k_n = L^{-1}\lambda^n$ (with $L^{-1}=0.05$
being the wave number associated with the integral scale, and
$\lambda=2$ defining the ratio between successive shells). The
nonlinear energy transfer is characterized by sudden bursts of
activity (typically referred to as
``instantons'')~\cite{biferale2003shell,daumont2000instanton}, where
anomalous fluctuations are spread from large- to small-scales.  The
complex space-time patterns and localized correlations in $v(t)$,
given by these intermittent bursts, make a 1-dimensional Convolutional
Neural Network a well-suited choice for our neural network model (see
Methods and SI for details).

\begin{figure}
   \centering \includegraphics[width=.49\linewidth]{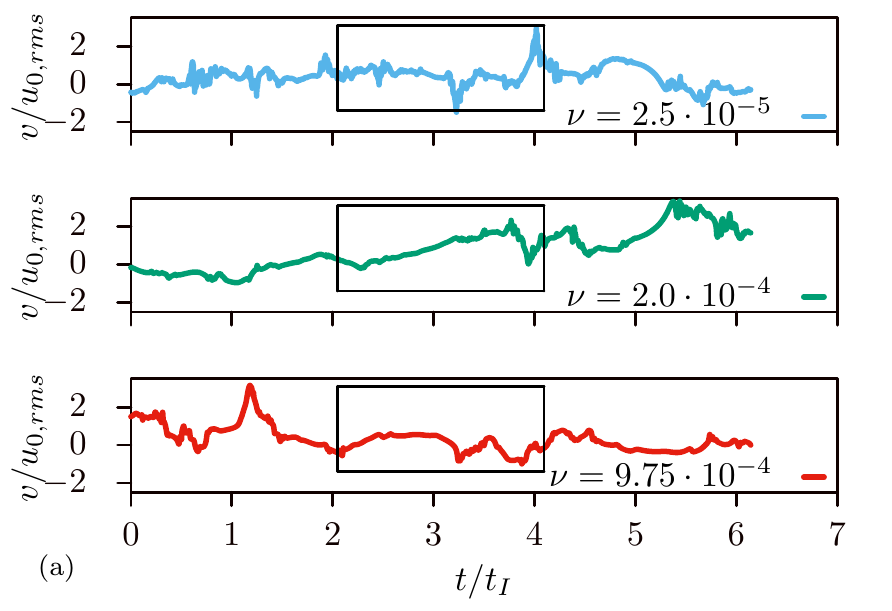}      \includegraphics[width=.49\linewidth]{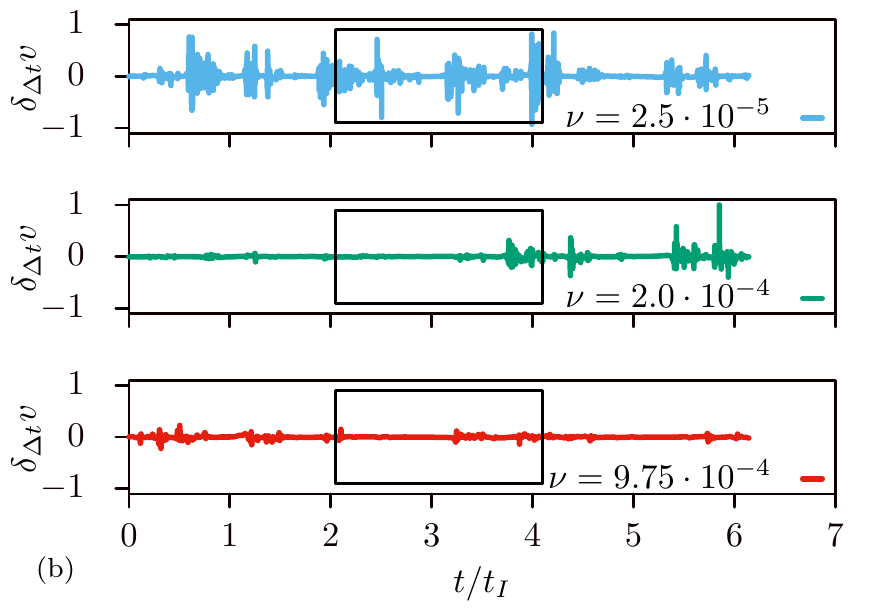}   
   \caption{%
     Velocity signals, $v(t)$, used to train the DNN for three
     different values of the viscosity $\nu$. The signals are
     normalized with the rms of the integral-scale velocity,
     $u_{0,rms}\approx v_{rms}$. The time is reported in units of eddy
     turnover times of the integral scale, $T_L$
     ($T_L\approx 1000\Delta t$, being $\Delta t$ the time sampling of
     the signals input to the DNN).  Each training signal spans $2048$
     samples, i.e. about two eddy turnover times (the rectangular
     frames identify individual training signals). (b) Velocity
     increments, $\delta_{\Delta t} v(t) = v(t + \Delta t)-v(t)$, computed with
     time interval $\Delta t$. Lower viscosity values yield higher
     turbulence intensity, thus more intermittent high-frequency
     components and more intense small-scale velocity differences.
   }
  \label{fig:timehistory}
\end{figure}

We train the DNN using a collection of datasets corresponding to
different viscosity values for our Lagrangian turbulent signal. Each
dataset includes a large number of velocity signals (few
thousands) sampled over $2048$ time instants (see examples in
Figure~\ref{fig:timehistory}(a)). We decided to employ an external
forcing to maintain the root-mean-square energy fluctuations of the
signals, and thus $v_{rms}$, statistically stationary. As a result,
the viscosity fully determined the turbulence intensity, and therefore
the Reynolds number. Decreasing the viscosity increases the
high-frequency content of the velocity signals (as~$\sim Re^{1/2}$,
cf. time-increments in Figure~\ref{fig:timehistory}(b))
by reducing the dissipative time- and length-scales. The resulting
wider inertial range reflects the higher turbulence intensity. We train
the network in a supervised way to infer the viscosity from the
velocity signals; the collection of datasets covered uniformly the
viscosity interval $10^{-5}\leq \nu \leq 10^{-3}$ in $39$ equi-spaced
levels.

\begin{figure}
  \centering
  \includegraphics[width=.80\linewidth]{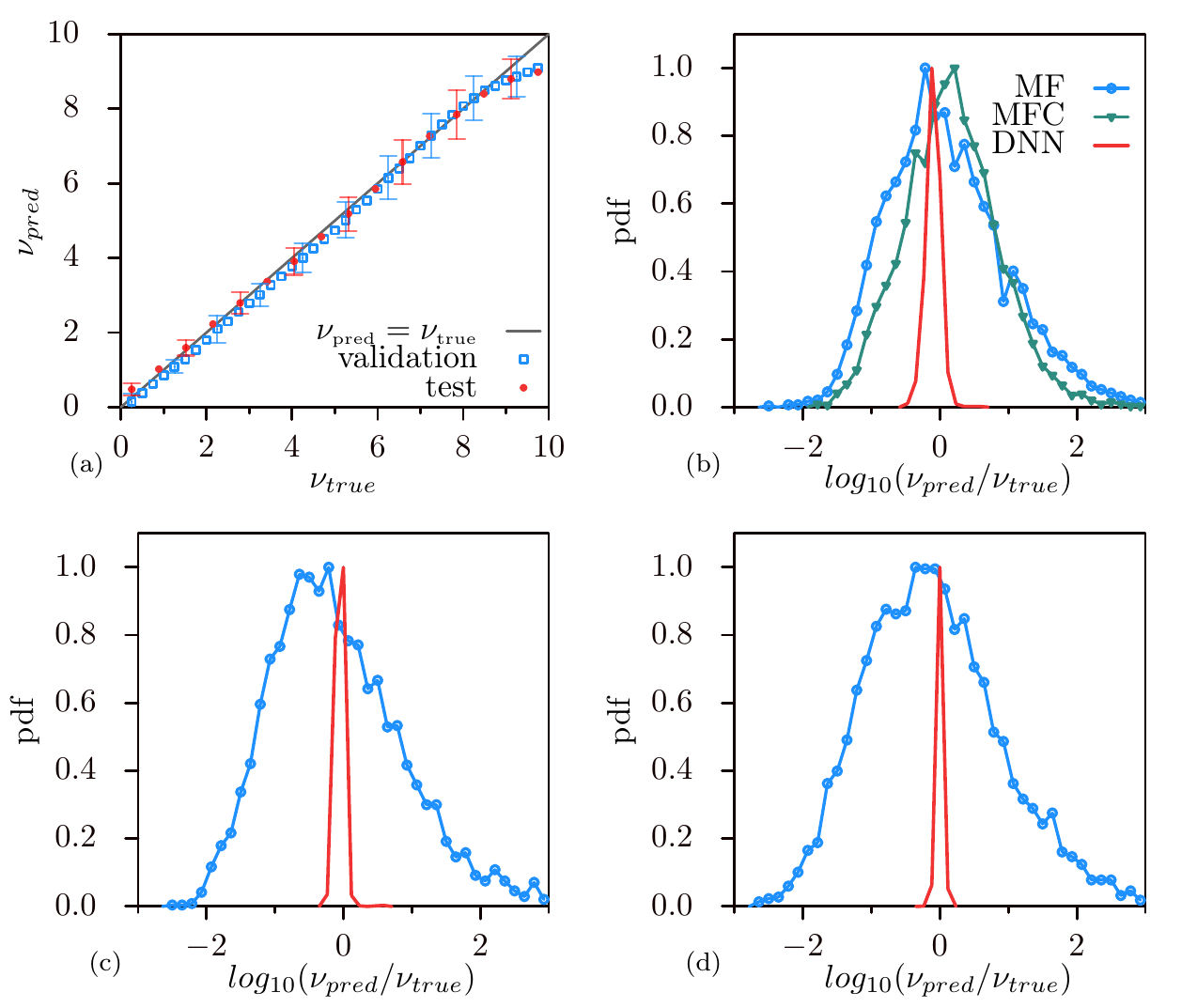}
  \caption{%
    (a) Average predictions of viscosity, $\nu$, by the DNN
    ($\nu_{pred}$, $y$-axis) vs. ground truth ($\nu_{true}$,
    $x$-axis), for the validation and test sets considered (the axes
    are scaled by a factor $10^{-4}$).  The
    diagonal line identifies error-free predictions, i.e.
    $\nu_{pred} = \nu_{true}$.  We include a few indicative error bars
    of size $\pm \sigma$ from the average, to indicate the typical
    spread of the prediction.  (b,c,d) Comparison of the viscosity
    estimates over three viscosity levels in the validation set,
    respectively, $\nu=0.000075$, $0.0002$ and $0.0007$). We report
    the pdf of $\log_{10}(\nu_{pred}/\nu_{true})$ for the DNN (solid
    line), and for the multifractal model (dotted-line),
    Eq.~(\ref{1}).  Evaluating $v_{rms}$ in Eq.~(\ref{1}) through an
    ensemble averaged (MF), or individually for each signal as
    $v_{rms} = \frac{1}{2} S^2(\infty) \approx \frac{1}{2}S^2(T)$
    (MFC), yields similar results, which we report panel (b). We
    notice how the predictions based on Eq.~(\ref{1}), once normalized
    to the true value, range within about four orders of
    magnitude, whereas they remain confined within $15\%$ accuracy
    in the case of the DNN.
  }\label{fig:compareMF}
\end{figure}

\begin{figure}
  \centering
  \includegraphics[width=.50\linewidth]{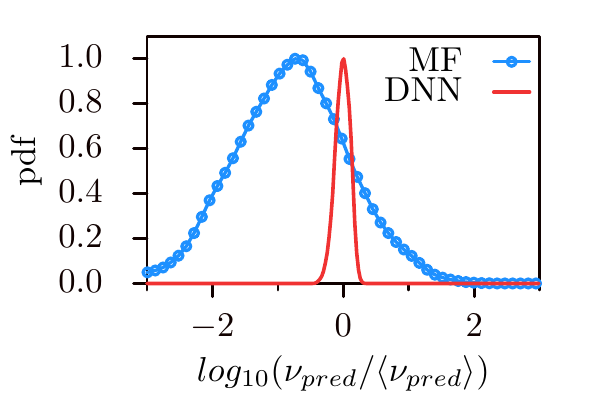}
  \caption{%
    Comparison of the viscosity estimates by the DNN (trained through
    shell model data, solid line), and by the multifractal model (MF,
    dotted-line) for Lagrangian velocity signals obtained by a DNS
    simulation. We normalize the estimates by the most frequent
    prediction, $\langle\nu_{pred}\rangle$, i.e. we report the pdf of
    $\log_{10}(\nu_{pred}/\langle\nu_{pred}\rangle)$. This enables a
    comparison of the prediction root-mean-squared errors. As in the
    considered validation and test cases, the DNN estimates fall
    within a significantly smaller range than in case of predictions
    by the multifractal model. Note that a log-normal distribution
    is expected for the viscosity
    estimates~\cite{benzi1985characterisation}, thus the discrepancy
    between average and most frequent prediction, clearly evident in
    the MF case.   %
  }
  \label{fig:compareDNS}
\end{figure}

We assess the predictive performance of the DNN by considering unseen
signals generated by the shell model. Results are reported in
Figure~\ref{fig:compareMF} where two different test sets are used: 1)
``validation set'': including statistically independent realizations
of the velocity signals than the training phase, yet having the same
viscosity values; 2) ``test set'': signals having different viscosity
values than considered during training, yet within the same viscosity
range. The results demonstrate that the network is capable of very
accurate predictions for the viscosity over the full range considered,
also for viscosity values different from those employed in training.
By aggregating the viscosity estimates over a large number of
statistically independent shell model signals with fixed viscosity, we
can define an average estimate as well as a root-mean-squared error,
see Figure~\ref{fig:compareMF}(a).

To further reflect on this remarkable result, we turn to a
physical argument for estimating the viscosity. For the second-order
Lagrangian structure functions
$S^2(\tau) = \langle\delta_\tau v(t)^2\rangle$, where
$\delta_\tau v(t) = v(t+\tau) - v(t)$, in steady HIT the following
estimate holds in the small $\tau$ limit: 
\begin{equation}
S^2(\tau)  = A v_{rms}^2 \left(\frac{\tau}{T_L}\right)^2Re^\alpha,
\label{1}
\end{equation}
where $A$ is a constant of order $1$, $\alpha = 0.57$, and
$v_{rms}^2/T_L \equiv \epsilon$ is the Reynolds-independent rate of
energy dissipation~\cite{arneodo2008universal}.  The (dissipative)
time-scale $\tau_\eta$ is defined by the relation
$S^2(\tau_\eta)/\tau_\eta = \epsilon$ which, using Eq.~(\ref{1}),
gives $\tau_\eta/T_L = Re^{-\alpha}/A$.  Assuming $L$ and $v_{rms}$
known, or, alternatively estimating
$v_{rms}^2 = 1/2 S^2(\infty)\approx 1/2 S^2(2\,T_L)$ on a
signal-by-signal basis, to amend for large-scale oscillations within
the observation window, we evaluate $A$ by using a reference case of
known viscosity, say $\nu_0$. Figure~\ref{fig:compareMF}(b,c,d)
compare the pdfs of the logarithmic ratio between the estimated and
true viscosity, for estimates by the DNN and based on Eq.~(\ref{1}).
Predictions in case of Eq.~(\ref{1}) spread over a range
$\nu_{pred}/\nu_{true}\in[10^{-2},10^2]$, whereas this range is just
of order $15\%$ in case of the DNN. Besides, in
Figure~\ref{fig:compareMF}(b), we observe that evaluating $v_{rms}$ on
a signal-by-signal basis reduces, yet minimally, the variance of
viscosity predictions based on Eq.~(\ref{1}).

The high variance and heavy tails of the pdf of viscosity estimates
produced by Eq.~(\ref{1}) follow from the very limited statistical
sampling ($2048$ points), which is severely affected by large-scale
oscillations and small-scale intermittent fluctuations.  Because of
these, statistical convergence and, therefore, a stable value for the
LHS of Eq.~(\ref{1}), are attained only after very long observation
times.

\bigskip

Our DNN model can be tested on real Lagrangian velocity signals,
$v(t)$, obtained by the numerical integration of the Lagrangian
dynamics of a tracer particle in HIT, see
Figure~\ref{fig:compareDNS}. The underlying Eulerian velocity field is
obtained from Direct Numerical Simulation~\cite{bec2010intermittency}
of the Navier-Stokes equation at $Re_{\lambda}=400$ (see Methods).
Remarkably, the DNN, although trained on shell model data, is able to
estimate with extremely high accuracy the viscosity $\nu$ even in case
of real Lagrangian data (note that Lagrangian velocity signals from
DNS has been exhaustively validated against experimental data in the
past~\cite{toschi2009lagrangian}).  This points to the fact that the
DNN relies on space-time features that are equally present in the
shell model as well as in the real Lagrangian signals.

What is the best result that can be achieved according to the current
understanding of the physics of turbulence?  Both by direct estimation
of the Reynolds number or by viscous scale fitting, i.e. using
Eq.~(\ref{1}), the statistical accuracy is limited by the fluctuations
of the large-scale velocity. Therefore, the statistical error is
limited by the number of large-scale eddy turnover times.  As shown in
Figure~\ref{fig:compareMF}(b,c,d), a traditional statistical physics
approach produces estimates for the viscosity spread over four orders
of magnitude, while the DNN is capable of delivering accurate 
predictions, scattering within a $15\%$ range.

This points at two major results: first, the DNN, at least within the
range of the training signals, must be able to identify space-time
structures that strongly correlate with turbulence intensity and which
are rather insensitive to the strong fluctuations of the instantaneous
value of the large-scale velocity (cf. SI for a discussion). This
finding unlocks the possibility of defining, practically
instantaneously, turbulence intensities, Reynolds numbers, or
connected statistical quantities for complex flows and
fluids. Estimating locally, in space and in time, the turbulence
intensity at laminar-turbulent interfaces or from atmospheric
anemometric readings can now be possible. The quantitative definition
of an effective viscosity within boundary layers or complex fluids,
such as emulsions or viscoelastic flows, may similarly be
pursued. Finally, being able to extract the space-time correlations
identified by the DNN may give us novel and fundamental insights in
turbulence physics and the complex skeleton of the fluctuating
turbulent energy cascades.



\section*{Acknowledgments}
 The authors acknowledge the help of Pinaki Kumar for the
  development of the vectorized GPU code.

\bibliography{master}

\section*{Methods}

\subsection*{Generating the database of turbulent velocity signals}  We
employ the SABRA~\cite{l1998improved} shell model of turbulence to
generate Lagrangian velocity signals
$v(t) = \sum_{n \geq 0} \Re{u_n(t)}$ corresponding to different
turbulence levels (Reynolds numbers).  Shell models evolve in time,
$t>0$, the complex amplitude of velocity fluctuations, $u_n(t)$, at
logarithmically spaced wavelengths, $k_n = k_0\lambda^n$
($n=0,1,\ldots$).

The amplitudes  $u_n(t)$ evolve according to the following equation:
\begin{eqnarray}
\frac{du_n(t)}{dt} = i ( ak_{n+1}u_{n+2}u^*_{n+1} +bk_n
u_{n+1}u^*_{n-1} - c k_{n-1}u_{n-1}u_{n-2} ) - \nu k_n^2 + f_n(t),
\label{shellm}
\end{eqnarray}
where $\nu > 0$ represents the viscosity, $f_n(t)$ is the forcing, and
the real coefficients $a,b,c$ regulate the energy exchange between
neighboring shells. We consider the following constraints: $a+b+c=0$,
which guarantees conservation of energy $E = \sum_{n} |u_n|^2$, for an
unforced and inviscid system ($f_n = 0$, $\nu = 0$, respectively);
$b=-1/2$, which gives to the second (inviscid/unforced) quadratic
invariant of the system, $H = \sum_{n\geq 0} (-)^n k_n |u_n|^2 $, the
dimensions of an helicity; to fix the third parameter we opt for the
common choice $c = 1/2$.  We truncate Eq.~(\ref{shellm}) to a finite
number of shells $0\leq n <\ N = 28$ which ensures a full resolution
of the dissipative scales in combination with our forcing and
viscosity range.  We simulate the system in Eq.~(\ref{shellm}) via a
$4^{th}$ Runge-Kutta scheme with viscosity explicitly
integrated~\cite{bohr2005dynamical} (the integration step, $dt$, is
fixed for all simulations, to be about three orders of magnitude
smaller than the dissipative time-scale for the lowest viscosity
case).

We inject energy through a large-scale forcing acting on the
first two shells~\cite{l1998improved}.  The forcing dynamics is given
by an Ornstein-Uhlenbeck process with a timescale matching the eddy
turnover of the forced shells ($\tau_n = (k_n u_n)$,
$n=0,1$). Additionally, we set the ratio
$|\sigma(f_0)/\sigma(f_1)| = \sqrt{2}$ between the standard deviation
($\sigma(f_n)$) of the two forcing signals. This ensures a
helicity-free energy flux in the system~\cite{l1998improved}.  See the
SI for further information on the signals and values of the constants.

We generate the signals in a vectorized fashion on an nVidia V100
card.  We integrate simultaneously $15.000$ instances of the system in
Eq.~(\ref{shellm}) in a vectorized manner (i.e. system description by
$15.000\times 28$ complex variables), and dump the state $55.000$
times after skipping the first $5.000$ samples. 

\subsection*{Lagrangian velocity signals from Direct Numerical Simulations}
The true Lagrangian velocity signals are obtained from the numerical
integration of Lagrangian tracers dynamics evolved on top of a Direct
Numerical Simulation of HIT turbulence. The Eulerian flowfield is
evolved via a fully de-aliased algorithm with second-order
Adams-Bashfort time-stepping with viscosity explicitly integrated. The
Lagrangian dynamics is obtained via a tri-linear interpolation of the
Eulerian velocity field coupled with second-order Adams-Bashfort
integration in time. The Eulerian simulation has a resolution of
$2048^3$ grid points, a viscosity of $3.5\cdot 10^{-4}$, a timestep
$dt=1.2\cdot 10^{-4}$, this corresponded to a $Re_{\lambda}~ 400$,
dissipative scale $\eta =3\cdot 10^{-3}$ and
$\tau_{\eta} = 2\cdot 10^{-2}$. The Lagrangian trajectories employed
are available at the 4TU.Centre for Research Data~\cite{LagrDNSDATA}.

\subsection*{Deep Neural Network (DNN)}
We employ a one-dimensional Convolutional Neural Network (CNN)
architecturally inspired by the VGG
model~\cite{simonyan2014very}. Developing a neural network model poses
the major challenge of selecting a large number of
hyperparameters. This particular architecture deals with this issue by
fixing the size of the filters and employs stacks of convolutional
layers to achieve complex detectors.  For our model we opted for
convolutional filters of size $3$, which is comparable or smaller than
the dissipative time-scale of the turbulent signals.  The network
includes four blocks, each formed by three convolutional layers
(including $128$ filters each), a max pooling layer (window: 2) and a
dropout layer, that capture all the spatial features of the signal
(cf. DNN architecture in SI). These layers are followed by a
fully-connected layer with $128$ neurons and Re-Lu activation that
collects all the spatial features into a dense representation. The
final layer provides a linear map from the dense representation to the
estimated viscosity.  A complete sketch of the network is in the SI.

\subsection*{DNN Training} We train the neural network in a supervised fashion
and with $L^2$ training loss to output a continuous value in the
interval $[-1,1]$, which is linearly mapped to $[\min\nu,\max\nu]$.
The training set is composed of $192.000$ turbulent velocity signals
(time-sampled over $2048$ points) uniformly distributed among $39$
viscosity levels (training-validation ratio: 75\%-25\%). See Table in
SI for further information.


\newcommand{\figOne}{Figure~1}
\newcommand{\figTwo}{Figure~2}
\newcommand{\figThree}{Figure~3}
\newcommand{\EqOne}{Eq.~(1)}
\newcommand{\EqTwo}{Eq.~(2)}

\section*{Supplementary Information (SI)}

\renewcommand{\thefigure}{S.\arabic{figure}}
\setcounter{figure}{0}

\subsection*{Width of the inertial range}
\begin{figure}[ht]
\centering
\includegraphics[width=.45\linewidth]{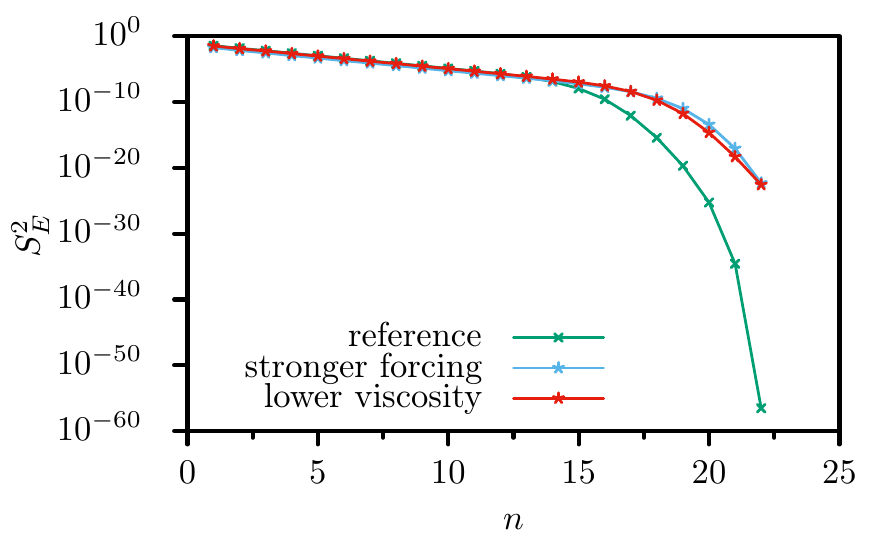}
\caption{Impact of the variation of forcing (operating at the large
  scale, $L$) or viscosity (regulating the small scale, $\eta$) on
  Eulerian structure functions. We compare a reference case with,
  respectively, a dynamics characterized by increased forcing
  (structure function translated and superimposed, \textit{a
    posteriori}, to the reference), and a dynamics characterized by
  decreased viscosity. Both these two cases yield a higher Reynolds
  number wider extension of the inertial range. 
  \label{fig:comparison} }
\end{figure}

In presence of limited statistics as in the case of relatively short
signals, the estimation of the width of the inertial range or,
similarly, the estimation of the viscosity, is enslaved to large scale
energy fluctuations. On a time scale comparable to the large scale
fluctuations, local increments or decrements of the system energy
yield almost instantaneous widenings or shortenings of the inertial
range. This effect can be naturally interpreted in terms of viscosity,
where local energy increments play the same effect of a lower
viscosity on the width of inertial range (see
Figure~\ref{fig:comparison}, where show this aspect for Eulerian
structure functions).

\begin{figure}
  \centering
  \includegraphics[width=.45\linewidth]{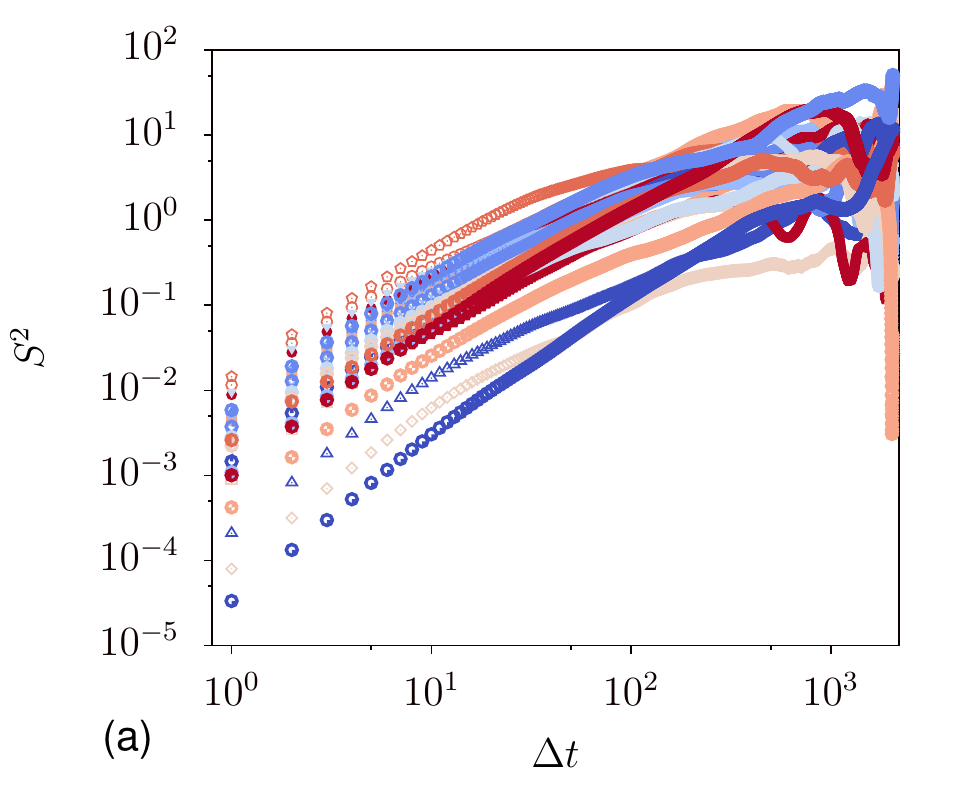}
  \includegraphics[width=.45\linewidth]{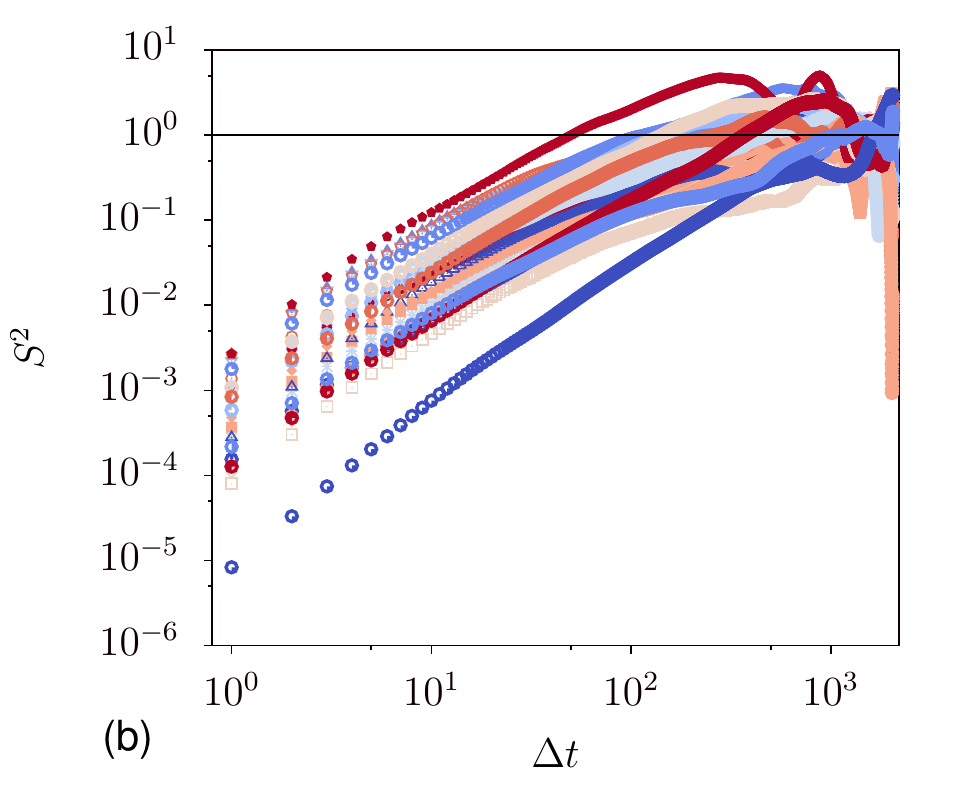}
  \caption{Collection of second order Lagrangian structure functions,
    $S^2$,  without (a) and with (b) normalization with respect to
    the integral scale energy, i.e. the asymptotic value
    $S^2(\infty) = 2 v_{rms}^2 \approx S^2(T)$. Each
    plot reports a collection of $25$ structure functions extracted
    from the training set and with associated viscosity
    $\nu=0.0005$. The x-axis is in units of sampling time, $\Delta t$,
    as presented to the DNN.}
  \label{fig:Lagrf}
\end{figure}

In Figure~\ref{fig:Lagrf}(a), we report Lagrangian structure functions
for a set of training signals with fixed viscosity values. The limited
statistics yield high fluctuations among the structure function, due
to a combination of large-scale energy fluctuations and small scale
intermittency. In Figure~\ref{fig:Lagrf}(b), we amend large-scale
fluctuations by normalizing by the signal energy, i.e. we report
$S^2(\tau)/S^2(\infty)$.

\subsection*{Data generation, training, testing and neural network
  parameters}
\begin{figure}[th]
  \centering
  \includegraphics[width=.32\linewidth]{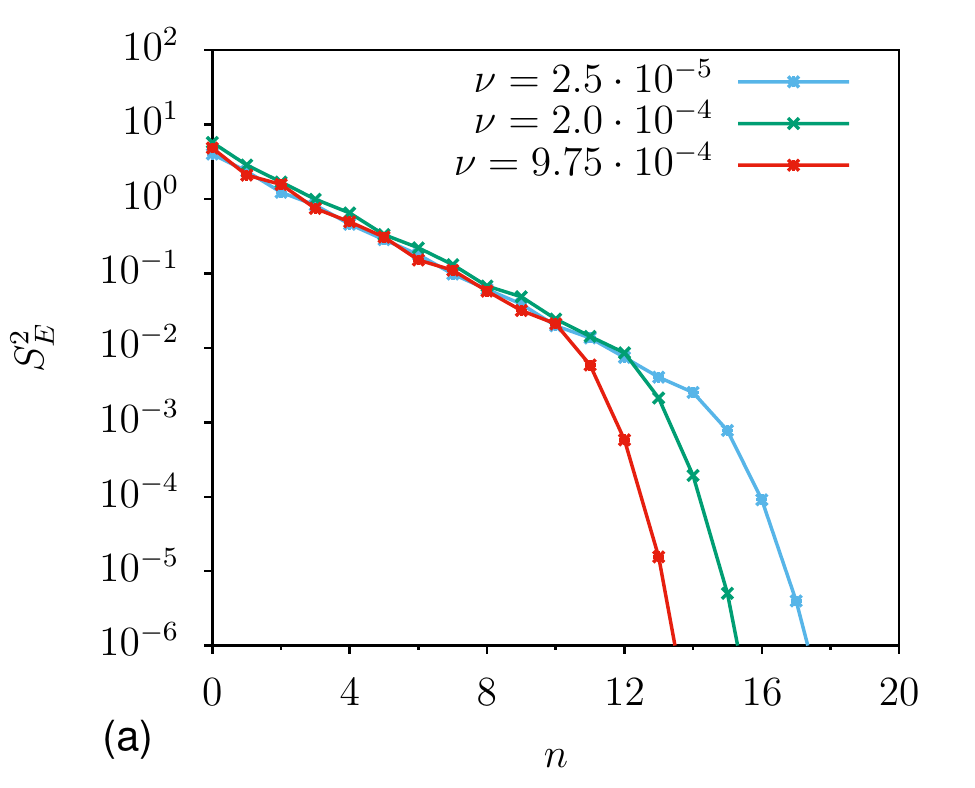}
  \includegraphics[width=.32\linewidth]{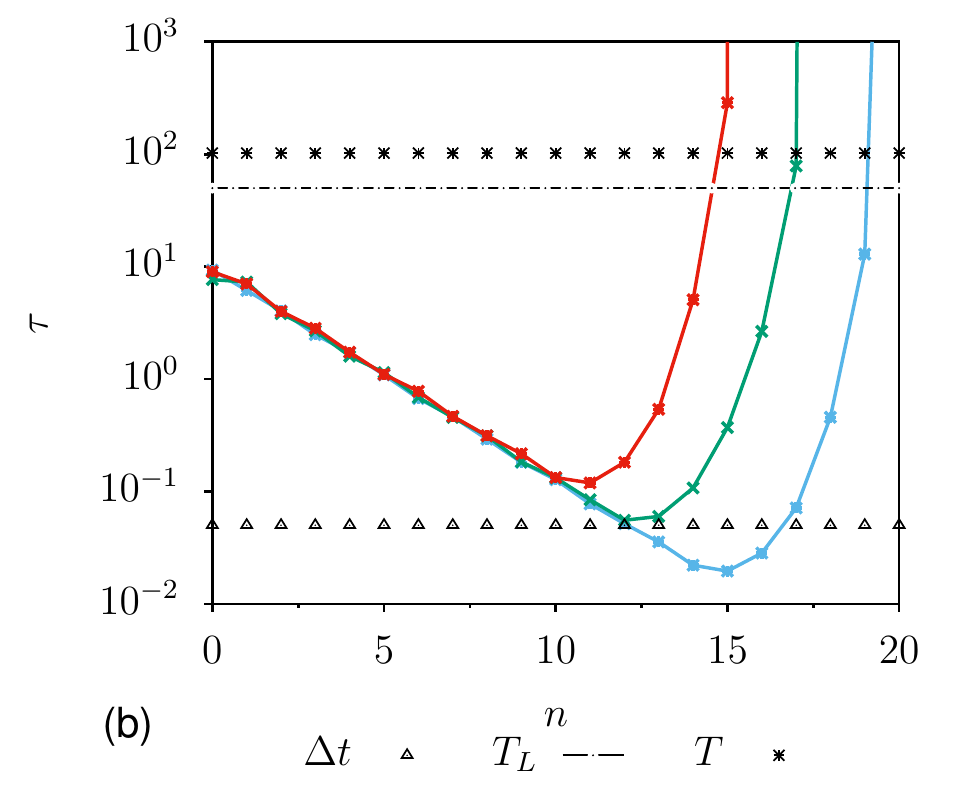}  \includegraphics[width=.32\linewidth]{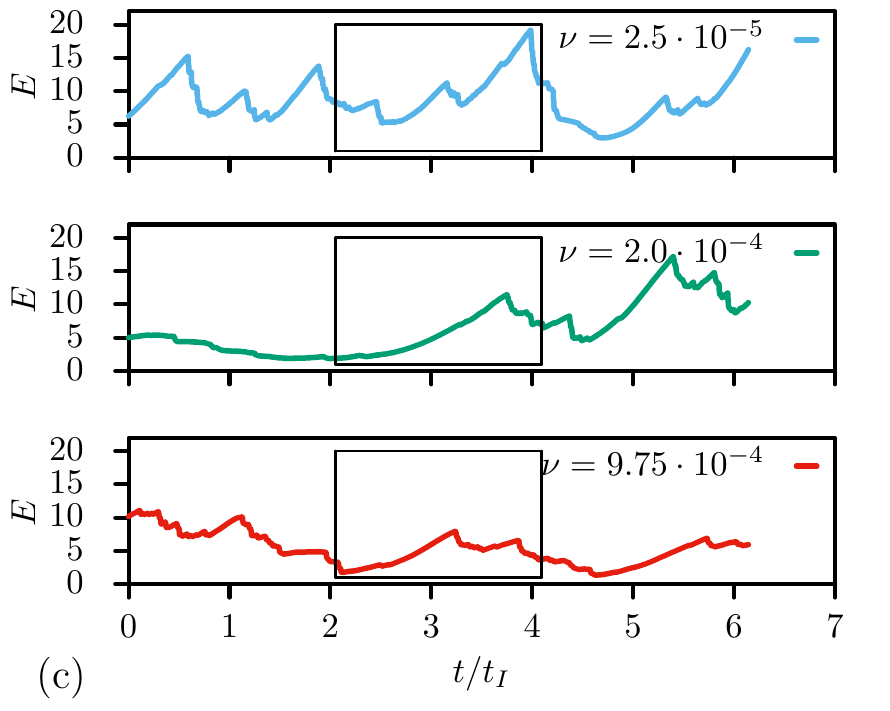}
  \caption{(a) Eulerian structure functions
    $S^2_E(n) = \langle|u_n|^2\rangle$ for the same three viscosity
    cases reported in \figOne. Reducing the viscosity leads to an
    extension of the inertial range, while the energy content of the
    larger scales remains unchanged.  (b) A scale-by-scale estimate of
    the correlation times for the shell models (via the inertial
    scaling $\tau_n \sim (k_n u_{n,rms})^{-1}$) for three different
    viscosity values. The observation window $T$, the calculated
    decorrelation time of the integral scale $T_L$, and of the DNN
    sampling time $\Delta t$ are reported. (c) Energy
    time-histories for the signals reported in \figOne.}
  \label{fig:structuref}
\end{figure}

\begin{figure}[th]
  \centering
    \begin{tikzpicture}[x=1.5cm, y=1.5cm, >=stealth]
      \newcommand{\blockSmdist}{.4}
      \newcommand{\blockLdist}{.5}
      \newcommand{\locFsz}{\footnotesize}

      \coordinate (beg_sg) at   (0,0) ;

      \node(TT) [ anchor=center , rectangle, draw, fill=white]  at ([shift={(1.,.5)}]beg_sg) {\locFsz input. size: $(N,1) = (2048,1)$};

      \begin{axis}[
    at=(beg_sg),
    yticklabels={,,},
    xticklabels={,,},
    ticks=none,
    height=2cm,
    x=0.01cm,
    axis x line=bottom,axis y line=left 
  ]
    \addplot[color=black,mark=., 
        point meta=explicit symbolic] coordinates {
(0,0.682913) (1,0.756124) (2,0.822830) (3,0.842523) (4,0.918590) (5,0.988522) (6,1.050295) (7,1.083329) (8,1.067608) (9,1.043301) (10,1.038942) (11,1.049854) (12,1.071483) (13,1.106384) (14,1.161781) (15,1.237457) (16,1.286376) (17,1.229043) (18,1.151414) (19,1.117566) (20,1.124484) (21,1.158026) (22,1.126305) (23,1.016059) (24,1.020702) (25,1.065855) (26,1.108564) (27,1.100526) (28,1.025105) (29,1.039807) (30,1.069213) (31,1.078565) (32,1.071342) (33,1.044331) (34,0.997742) (35,0.949450) (36,0.949198) (37,1.061239) (38,0.951381) (39,0.651155) (40,0.532879) (41,0.820815) (42,0.852155) (43,0.828857) (44,0.819557) (45,0.812153) (46,0.798261) (47,0.767064) (48,0.707647) (49,0.632256) (50,0.634127) (51,0.675650) (52,0.651614) (53,0.610088) (54,0.559457) (55,0.504993) (56,0.468545) (57,0.474563) (58,0.507537) (59,0.539402) (60,0.566411) (61,0.590910) (62,0.614230) (63,0.637191) (64,0.660473) (65,0.684740) (66,0.710660) (67,0.738822) (68,0.769787) (69,0.804056) (70,0.842108) (71,0.884252) (72,0.930698) (73,0.981537) (74,1.036708) (75,1.096017) (76,1.159198) (77,1.225856) (78,1.295425) (79,1.366933) (80,1.438851) (81,1.508640) (82,1.572553) (83,1.626000) (84,1.666961) (85,1.711066) (86,1.816606) (87,1.810080) (88,1.631830) (89,1.648950) (90,1.369685) (91,1.410165) (92,1.534281) (93,1.602501) (94,1.616883) (95,1.605341) (96,1.565141) (97,1.482688) (98,1.351002) (99,1.260594) (100,1.288641) (101,1.247071) (102,1.258869) (103,1.165300) (104,1.156616) (105,1.225455) (106,1.191490) (107,1.150772) (108,1.130795) (109,1.099123) (110,1.052573) (111,0.998860) (112,0.952741) (113,0.915138) (114,0.914682) (115,0.944282) (116,0.963168) (117,0.970660) (118,0.966867) (119,0.951915) (120,0.927917) (121,0.901918) (122,0.888283) (123,0.898585) (124,0.919349) (125,0.930450) (126,0.931036) (127,0.926137) (128,0.919580) (129,0.913933) (130,0.910933) (131,0.911874) (132,0.917445) (133,0.926356) (134,0.931229) (135,0.908513) (136,0.804407) (137,0.617815) (138,0.814148) (139,0.905984) (140,0.927059) (141,0.982394) (142,1.097300) (143,1.171975) (144,1.262813) (145,1.022084) (146,0.912536) (147,0.890375) (148,0.871024) (149,0.852628) (150,0.868979) (151,0.965091) (152,1.026326) (153,1.068987) (154,1.136146) (155,1.237343) (156,1.225271) (157,1.132057) (158,1.193692) (159,1.274905) (160,1.460079) (161,1.422544) (162,1.313453) (163,1.296343) (164,1.282656) (165,1.240106) (166,1.169858) (167,1.176269) (168,1.310166) (169,1.379914) (170,1.412404) (171,1.421999) (172,1.414374) (173,1.429279) (174,1.542867) (175,1.614703) (176,1.607766) (177,1.598013) (178,1.584147) (179,1.554472) (180,1.505886) (181,1.469516) (182,1.476116) (183,1.486935) (184,1.499072) (185,1.515017) (186,1.533151) (187,1.552514) (188,1.573057) (189,1.595065) (190,1.618979) (191,1.645148) (192,1.673576) (193,1.703283) (194,1.730270) (195,1.744046) (196,1.728255) (197,1.679299) (198,1.620375) (199,1.568417) (200,1.519868) (201,1.468573) (202,1.416168) (203,1.381664) (204,1.412695) (205,1.529285) (206,1.608294) (207,1.609799) (208,1.590193) (209,1.572168) (210,1.560083) (211,1.555398) (212,1.558807) (213,1.569451) (214,1.584219) (215,1.598655) (216,1.609186) (217,1.614708) (218,1.616168) (219,1.615619) (220,1.615839) (221,1.620471) (222,1.634266) (223,1.662339) (224,1.705403) (225,1.744374) (226,1.719666) (227,1.620827) (228,1.580220) (229,1.615294) (230,1.676417) (231,1.749029) (232,1.752031) (233,1.586888) (234,1.502378) (235,1.473079) (236,1.462231) (237,1.456492) (238,1.450805) (239,1.443581) (240,1.435382) (241,1.429055) (242,1.429619) (243,1.441905) (244,1.466097) (245,1.496241) (246,1.525113) (247,1.548818) (248,1.566595) (249,1.578961) (250,1.586609) (251,1.590059) (252,1.589619) (253,1.585463) (254,1.577774) (255,1.566816) (256,1.553046) (257,1.537409) (258,1.521581) (259,1.508393) (260,1.502026) (261,1.506711) (262,1.519345) (263,1.506871) (264,1.417689) (265,1.356836) (266,1.367979) (267,1.391909) (268,1.416404) (269,1.439944) (270,1.456533) (271,1.454573) (272,1.431197) (273,1.407638) (274,1.402483) (275,1.411653) (276,1.426147) (277,1.441347) (278,1.455951) (279,1.469835) (280,1.483156) (281,1.495934) (282,1.508099) (283,1.519471) (284,1.529832) (285,1.539053) (286,1.547191) (287,1.554412) (288,1.561051) (289,1.567461) (290,1.573957) (291,1.580742) (292,1.587909) (293,1.595526) (294,1.603608) (295,1.612182) (296,1.621182) (297,1.630599) (298,1.640411) (299,1.650550) (300,1.660996) (301,1.671732) (302,1.682721) (303,1.693940) (304,1.705352) (305,1.716896) (306,1.728531) (307,1.740232) (308,1.751953) (309,1.763696) (310,1.775468) (311,1.787256) (312,1.799080) (313,1.810984) (314,1.823027) (315,1.835273) (316,1.847836) (317,1.860840) (318,1.874470) (319,1.888963) (320,1.904616) (321,1.921897) (322,1.941468) (323,1.963994) (324,1.989986) (325,2.019092) (326,2.049469) (327,2.077797) (328,2.100962) (329,2.117963) (330,2.130009) (331,2.138671) (332,2.144484) (333,2.146060) (334,2.139451) (335,2.117539) (336,2.070748) (337,1.997228) (338,1.935324) (339,1.951434) (340,2.004076) (341,2.033526) (342,2.040149) (343,2.033589) (344,2.019517) (345,2.000949) (346,1.980537) (347,1.961832) (348,1.948893) (349,1.944106) (350,1.946419) (351,1.952495) (352,1.959228) (353,1.964985) (354,1.969319) (355,1.972318) (356,1.974146) (357,1.974887) (358,1.974491) (359,1.972929) (360,1.970069) (361,1.965774) (362,1.959857) (363,1.952126) (364,1.942349) (365,1.930259) (366,1.915578) (367,1.897987) (368,1.877172) (369,1.852691) (370,1.824076) (371,1.790817) (372,1.752272) (373,1.707772) (374,1.656666) (375,1.598397) (376,1.532887) (377,1.461019) (378,1.385337) (379,1.310367) (380,1.241731) (381,1.183615) (382,1.136805) (383,1.099527) (384,1.070617) (385,1.052969) (386,1.057499) (387,1.105823) (388,1.194081) (389,1.182410) (390,1.117082) (391,1.078373) (392,1.036764) (393,0.991761) (394,0.945040) (395,0.897558) (396,0.849880) (397,0.802435) (398,0.755617) (399,0.709828) (400,0.665532) (401,0.623310) (402,0.583827) (403,0.547773) (404,0.515545) (405,0.486489) (406,0.457296) (407,0.419116) (408,0.354637) (409,0.259432) (410,0.290170) (411,-0.076865) (412,0.635843) (413,0.316063) (414,0.502934) (415,0.613691) (416,0.443546) (417,0.569949) (418,0.663181) (419,0.666062) (420,0.852852) (421,0.954834) (422,1.112511) (423,1.441916) (424,1.637267) (425,1.743654) (426,1.708569) (427,1.785759) (428,1.558707) (429,1.393465) (430,1.185370) (431,0.969970) (432,1.097956) (433,0.941884) (434,1.190832) (435,1.156925) (436,1.248844) (437,1.307548) (438,1.300322) (439,1.314186) (440,1.340297) (441,1.384908) (442,1.452605) (443,1.502858) (444,1.466999) (445,1.573940) (446,1.693970) (447,1.796635) (448,1.896846) (449,1.990605) (450,2.075609) (451,2.167427) (452,2.301065) (453,2.495360) (454,2.377648) (455,2.528710) (456,1.733119) (457,1.677180) (458,1.657370) (459,1.644836) (460,1.625472) (461,1.592949) (462,1.541397) (463,1.474532) (464,1.435435) (465,1.508821) (466,1.600450) (467,1.632528) (468,1.632813) (469,1.620152) (470,1.610461) (471,1.625026) (472,1.674499) (473,1.734666) (474,1.781783) (475,1.814612) (476,1.838991) (477,1.859667) (478,1.879530) (479,1.900235) (480,1.922713) (481,1.947439) (482,1.974344) (483,2.002802) (484,2.031635) (485,2.059527) (486,2.085420) (487,2.108958) (488,2.130405) (489,2.150418) (490,2.169608) (491,2.188268) (492,2.206276) (493,2.222956) (494,2.236940) (495,2.245838) (496,2.245759) (497,2.230975) (498,2.195188) (499,2.140041) 
    };
  \end{axis}
  \definecolor{fillConv}{HTML}{87CEFA}
  \definecolor{fillDens}{HTML}{F08080}

\node(CF) [ anchor=center,rectangle,rotate=-90,draw,fill=fillConv] at ([shift={(\blockSmdist,-1)}]beg_sg) {\locFsz conv $(f_s,f_M)$} ; 
\draw[ line width=.35mm] (TT) ++(-1.2,-.17) -- +(-.0,-1.) |- (CF); 
\draw[ line width=.35mm] (CF) -- +(8.16,0); 
\node(CF) [ anchor=center, rectangle,rotate=-90,draw,fill=fillConv
, drop shadow={top color=black,
              bottom color=white,
              shadow xshift=0.095em,
              shadow yshift=-0.095em,
              rounded corners }] at ([shift={(\blockSmdist,0)}]CF) {\locFsz conv $(f_s,f_M)$};
\node(CF) [ anchor=center, rectangle,rotate=-90,draw,fill=fillConv] at ([shift={(\blockSmdist,0)}]CF) {\locFsz conv $(f_s,f_M)$};
\node(CF) [ anchor=center, rectangle,rotate=-90,draw,fill=fillConv] at ([shift={(\blockSmdist,0)}]CF) {\locFsz max pool (2)};
\node(TT) [ anchor=center,rectangle,draw,dashed]  at ([shift={(0,-1.)}]CF) {\locFsz $(N/2,f_M)$};
\draw [->] (TT) -- (CF);
\node(CF) [ anchor=center, rectangle,rotate=-90,draw,fill=lightgray] at ([shift={(\blockSmdist,0)}]CF) {\locFsz dropout};

\node(CF) [ anchor=center,rectangle,rotate=-90,draw,fill=fillConv] at ([shift={(\blockLdist,0)}]CF) {\locFsz conv $(f_s,f_M)$} ; 
\node(CF) [ anchor=center, rectangle,rotate=-90,draw,fill=fillConv] at ([shift={(\blockSmdist,0)}]CF) {\locFsz conv $(f_s,f_M)$};
\node(CF) [ anchor=center, rectangle,rotate=-90,draw,fill=fillConv] at ([shift={(\blockSmdist,0)}]CF) {\locFsz conv $(f_s,f_M)$};
\node(CF) [ anchor=center, rectangle,rotate=-90,draw,fill=fillConv] at ([shift={(\blockSmdist,0)}]CF) {\locFsz max pool (2)};
\node(TT) [ anchor=center,rectangle,draw,dashed]  at ([shift={(0,-1.)}]CF) {\locFsz $(N/4,f_M)$};
\draw [->] (TT) -- (CF);
\node(CF) [ anchor=center, rectangle,rotate=-90,draw,fill=lightgray] at ([shift={(\blockSmdist,0)}]CF) {\locFsz dropout};

\node(CF) [ anchor=center,rectangle,rotate=-90,draw,fill=fillConv] at ([shift={(\blockLdist,0)}]CF) {\locFsz conv $(f_s,f_M)$} ; 
\node(CF) [ anchor=center, rectangle,rotate=-90,draw,fill=fillConv] at ([shift={(\blockSmdist,0)}]CF) {\locFsz conv $(f_s,f_M)$};
\node(CF) [ anchor=center, rectangle,rotate=-90,draw,fill=fillConv] at ([shift={(\blockSmdist,0)}]CF) {\locFsz conv $(f_s,f_M)$};
\node(CF) [ anchor=center, rectangle,rotate=-90,draw,fill=fillConv] at ([shift={(\blockSmdist,0)}]CF) {\locFsz max pool (2)};
\node(TT) [ anchor=center,rectangle,draw,dashed]  at ([shift={(0,-1.)}]CF) {\locFsz $(N/8,f_M)$};
\draw [->] (TT) -- (CF);
\node(CF) [ anchor=center, rectangle,rotate=-90,draw,fill=lightgray] at ([shift={(\blockSmdist,0)}]CF) {\locFsz dropout};

\node(CF) [ anchor=center,rectangle,rotate=-90,draw,fill=fillConv] at ([shift={(\blockLdist,0)}]CF) {\locFsz conv $(f_s,f_M)$} ; 
\node(CF) [ anchor=center, rectangle,rotate=-90,draw,fill=fillConv] at ([shift={(\blockSmdist,0)}]CF) {\locFsz conv $(f_s,f_M)$};
\node(CF) [ anchor=center, rectangle,rotate=-90,draw,fill=fillConv] at ([shift={(\blockSmdist,0)}]CF) {\locFsz conv $(f_s,f_M)$};
\node(CF) [ anchor=center, rectangle,rotate=-90,draw,fill=fillConv] at ([shift={(\blockSmdist,0)}]CF) {\locFsz max pool (2)};
\node(TT) [ anchor=center,rectangle,draw,dashed]  at ([shift={(0,-1.)}]CF) {\locFsz $(N/16,f_M)$};
\draw [->] (TT) -- (CF);
\node(CF) [ anchor=center, rectangle,rotate=-90,draw,fill=lightgray] at ([shift={(\blockSmdist,0)}]CF) {\locFsz dropout};

\node(FF1) [ anchor=center, rectangle,rotate=-90,draw] at ([shift={(\blockSmdist,-2.7)}]beg_sg) {\locFsz flatten};
\node(TT) [ anchor=center,rectangle,draw,dashed]  at ([shift={(0,-1.05)}]FF1) {\locFsz $(f_M N/16,1)$};
\draw [->] (TT) -- (FF1);

\draw[ line width=.35mm] (CF) -- (CF) ++(.25,0) -- ++(0,-1.3)  -- ++(-8.5,0)  |-  (FF1) ++(0,0); 

\node(FFE) [ anchor=east, rectangle,draw] at ([shift={(2.,0)}]FF1) {\locFsz output: $\nu$};

\draw[ ->, line width=.35mm] (FF1) -- (FFE); 

\node(FF) [ anchor=center, rectangle,rotate=-90,draw,fill=fillDens] at ([shift={(\blockLdist,0)}]FF1) {\locFsz dense};
\node(TT) [ anchor=center,rectangle,draw,dashed,fill=white]  at ([shift={(0,-.65)}]FF) {\locFsz $(f_d,1)$};

\draw [->] (TT) -- (FF);

\end{tikzpicture}
\caption{Feed-forward convolutional neural network considered. The
  network is constituted of four blocks each encompassing three
  convolutional layers (``conv'', filter size $f_s =3$, filter number
  $f_M = 128$, activation function: Re-Lu) one max pool layer that
  down scales the signal by a factor two and, in training, a dropout
  layer with dropout probability $20\%$. The dimensions of the feature
  map as obtained at the end of each block is reported in the dashed
  rectangles. The last feature map (dimension $(128,128)$), is densely
  connected to a representation layer which has $f_d =128$ dimensions
  and Re-Lu activation. The final output, i.e. the predicted viscosity
  $\nu$, is built from a linear combination of the dense
  representation values.  }
  \label{fig:NN}
\end{figure}
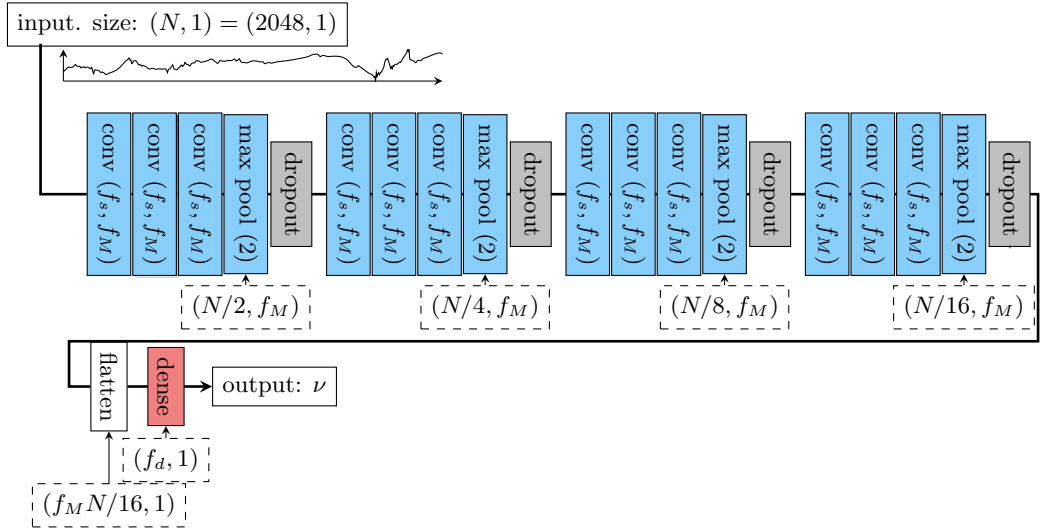

We include in Table~\ref{table:param} the parameters considered in the
shell model simulations by which the training, validation and test
datasets have been created. In Figure~\ref{fig:structuref} we
complement \figOne\ by including, for the same three viscosity levels,
further features of the considered signals. These are: (a) second
order Eulerian structure functions, $S^{2}_E(n)$ (where
$S^{p}_E(n)=S^p_E(k_n)=\left\langle |u_n|^p \right\rangle$) showing
that changing the viscosity only affects the extension of the inertial
range; (b) relevant time scales (computed by inertial scaling)
associated with the dynamics of the different shells; (c) signals
energy as a function of time. In Figure~\ref{fig:NN}, we report the
diagram of the neural network. Relevant structural parameters
(e.g. size of the convolutional filters) are reported in the figure
caption.

\begin{center}
  \begin{table}[ht!]
      \begin{tabular}{ l l l}
        Parameter & Value & \\
        \hline $N$ & $28$ & Number of shells
        \\ $k_0$ & $0.05$ & Wave number integral scale 
        \\ $\lambda$ & 2 & Inter-shell distance
        \\ $\sigma(f_0)$ & $2$ & Noise intensity forcing on shell $0$
        \\ $\sigma(f_1)$ & $2/\sqrt{2}$ & '' on shell $1$
        \\ $dt$ & $5\cdot 10^{-5}$ & integration step
        \\ $\Delta t$  & $1000\, dt$ & sampling time DNN
        \\ $T$   & $2048\, \Delta t$ & length window DNN
        \\
        \hline
      \end{tabular}      
      \begin{tabular}{ l l l }
        \hline
        Parameter & Training & Testing \\ \hline  
        $\min(\nu)$ & $2.5\cdot  10^{-5}$ & $6.0\cdot 10^{-5}$
        \\ $\max(\nu)$ & $9.75\cdot 10^{-4}$ & $9.6\cdot 10^{-4}$
        \\ increment $\nu$ & $2.5\cdot 10^{-5}$ & $6.0\cdot 10^{-5}$
        \\ levels & $39$ &  $16$ 
        \\ set size & 192.000 & 6.600
        \\ training:validation ratio & 75\%:25\% &N/A\\        
        \hline
        
      \end{tabular}
    \caption{(Top) Relevant parameters for the shell model its
      numerical integration; time length and sampling of the signals
      as provided to the deep neural network (DNN). (Bottom) Viscosity values
      considered for training validation and test; size of the related
      datasets.\label{table:param}}
\end{table}
\end{center}

\subsection*{Features observed by the DNN}
During training, the DNN develops feature detectors. As discussed in
the main text, we expect these detectors to select features
that, at the same time, strongly correlate with the turbulence
intensity and that are insensitive to large scale oscillations. As
generally expected in deep learning, detectors are likely
specific to the parameter range and statistical properties of the
signals contained in the training set.

In this section, to understand the characteristics of the signal that
our model relies on, we develop an ablation study by systematically
altering the content of randomly selected testing signals. The
modifications considered involve the suppression of frequency
components, or the random shuffling of the time structure. This
enables us to identify features mostly ignored by the DNN and,
conversely, restrict the set of characteristics of the signals
relevant for the DNN.

\begin{figure}[t]
  \centering
  \includegraphics[width=.46\linewidth]{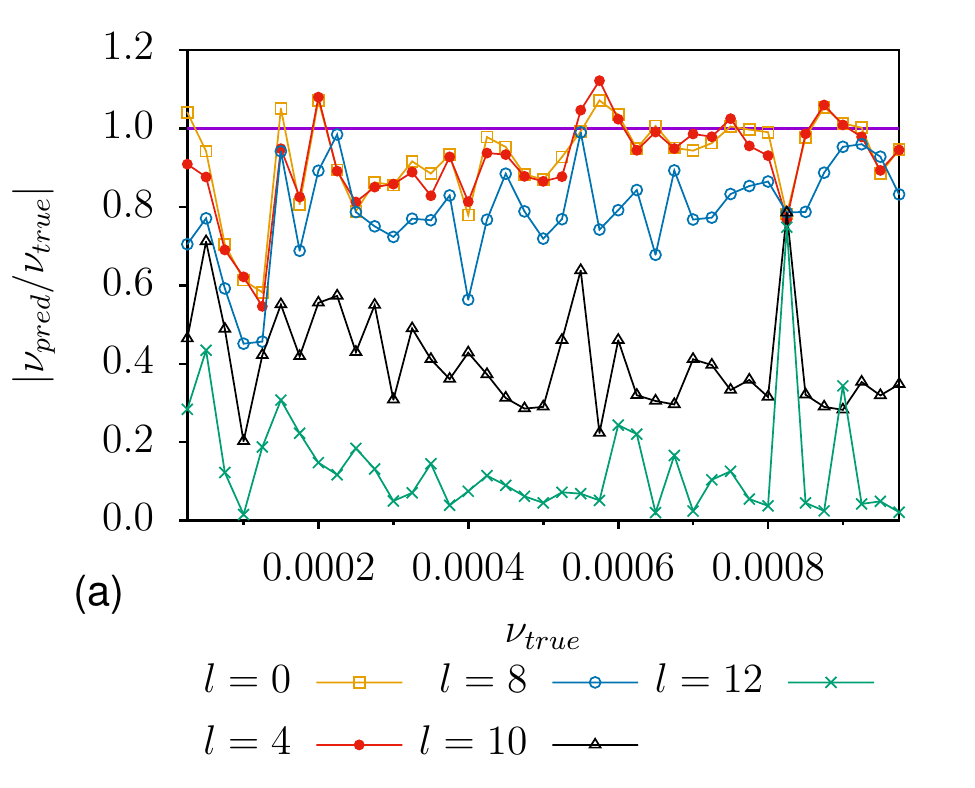}
  \includegraphics[width=.46\linewidth]{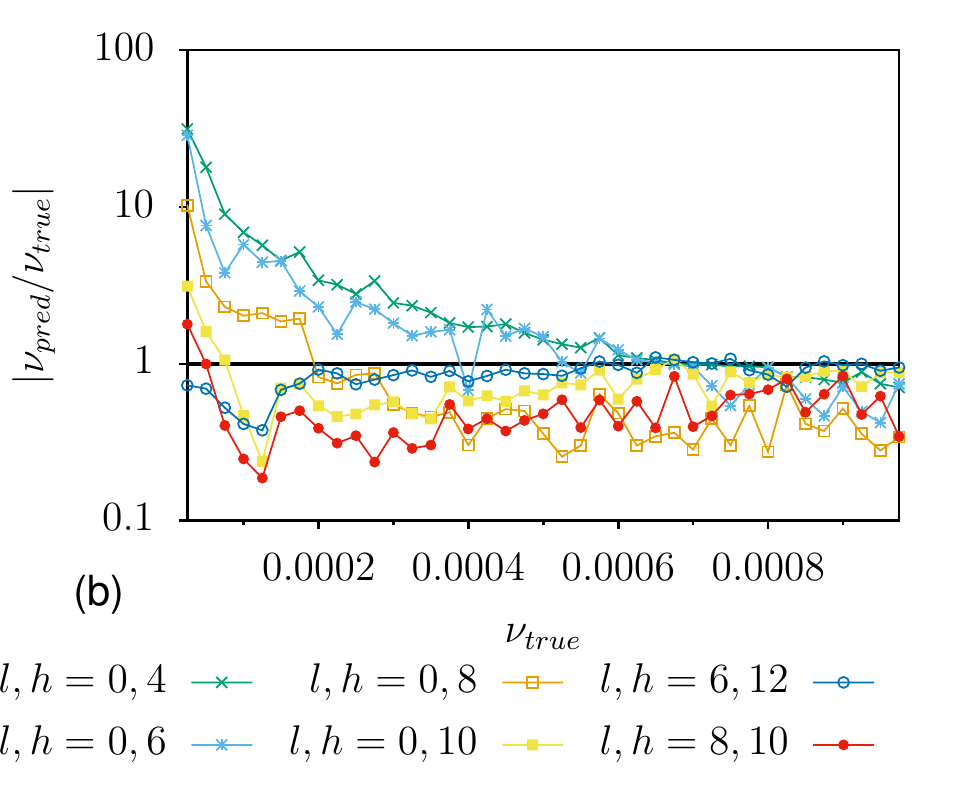}
  \caption{Viscosity predictions with ablated input signals from (a)
    an highpass filter: $v(t) = Re\sum_{n=l}^{26}u_n(t)$, (b) a
    bandpass filter: $v(t) = Re\sum_{n=l}^{h}u_n(t)$. In both
    cases, one single sample signal is considered for each viscosity
    value. Predictions are reported normalized with respect to the
    true value.}
  \label{fig:band-pass}
\end{figure}

In Figure~\ref{fig:band-pass} we consider testing signals that have
been altered through a high-pass (a) or a band-pass filter (b). In the
case of Lagrangian signals, filtering operations are easily performed
by restricting the summation in \EqTwo~to a subset of the shell
signals. We select one testing signal per viscosity level, we ablate
its spectral structure and we plot the DNN prediction. We notice that
the neural network is almost insensitive to the large scale dynamics,
as the estimates after the high-pass filter remain unaltered if the
large-scale shells are removed. We notice, in particular, that any
selection of a band of shells that includes the last part of the
inertial range yield almost error-free predictions.

\begin{figure}[h]
  \centering
  \includegraphics[width=.86\linewidth]{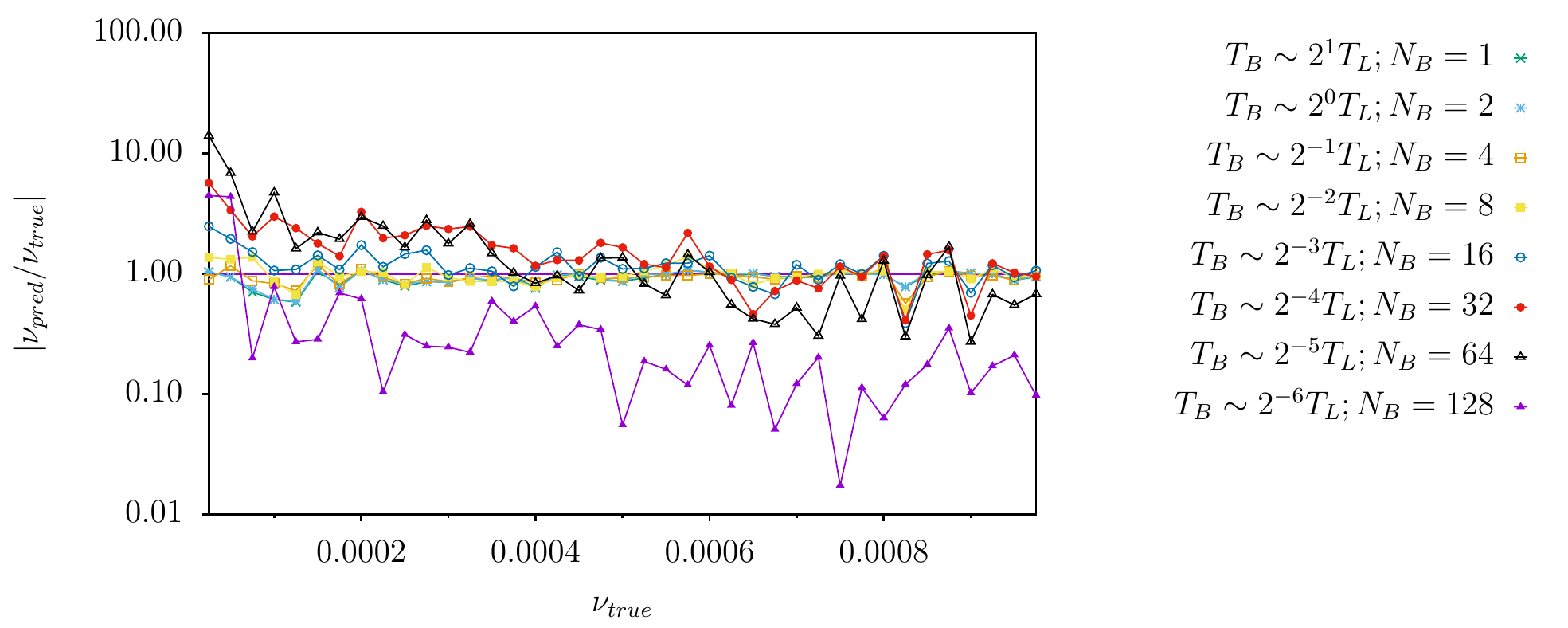}
  \caption{Viscosity predictions for block-based time-altered
    signals. Alteration is performed by splitting an initial signal in
    $N_B$ blocks (with time length $T_B$ reported in terms of the
    integral time scale) and then by performing a random permutation
    of the blocks.  One single sample signal is considered for each
    viscosity value. Predictions are reported normalized with the true
    value. }
  \label{fig:random:time:perm}
\end{figure}

Similarly, we can alter the time structure of the signals by
partitioning them in disjoint contiguous blocks of length $T_B$, and
then by randomly mixing these blocks. In
Figure~\ref{fig:random:time:perm} we report the predictions for
different block extensions. As the block extension remains in the same
order of the integral scale, the prediction remain mostly unaltered,
to then degrade as the block size become comparable to the dissipative
time-scale. This shows how the training develop feature extractors
targeting fine scales and correlations existing around the dissipative
end of the inertial range.

\end{document}